\documentclass{eas}
\usepackage{graphicx}
\usepackage{astron}
\begin{document}
\bibliographystyle{astron}
%%-----------------------------
%%      the top matter
%%-----------------------------
\title{A Search for Planets with SALT} 
\author{Monika Adam\'ow}\address{Toru\'n Centre for Astronomy, Nicolaus Copernicus University, ul. Gagarina 11, 87-100 Toru\'n; \email{adamow, aniedzi@astri.uni.torun.pl}}
\author{Andrzej Niedzielski}
\sameaddress{1}
\secondaddress{Department for Astronomy and Astrophysics, Pennsylvania State University, 525 Davey Laboratory, University Park, PA 16802}
\runningtitle{Adam\'ow \& Niedzielski: A Search for Planets with SALT}
\begin{abstract}
As the SALT High Resolution Spectrograph completion is nearing we plan to extend the Pennsylvania-Torun Planets Search (PTPS) with HET to the southern hemisphere. Due to  overlap of the skies available for both HET and SALT  in the declination range (+10, -10)  deg some cooperation and immediate follow up is possible.  Here we present,  as an example, a $\sim$ 1000 star  sample of evolved stars for the future  SALT Planet Search.

\end{abstract}
\maketitle
%%-----------------------------
%%      your text
%%-----------------------------
\section{The Pennsylvania - Torun  Planet Search  with HET}
Since the discovery of the first extrasolar planetary system, more than 300 planet candidates have been detected. Most of them were found by radial velocity measurements. Known host of extrasolar planets are generally main sequence stars. However, for better understanding of planetary systems formation and evolution it is necessary to extend existing planet searches to evolved and more massive stars.\\
The PTPS search for planet~\cite{niedz} is a  project of the Pennsylvania State University and Torun Centre for Astronomy  dedicated to search for planets around evolved stars using the radial velocity technique. This project is conducted with the 9.2m Hobby--Eberly Telescope~\cite{het}, located in Texas, equipped with the High Resolution Spectrograph~\cite{hrs}. The telescope is operated in the queue scheduled mode~\cite{qsm}. The spectrograph is used in the R=60000 resolution mode with a $I_2$ gas cell and it is fed with a 2 arcsec fiber. The basic data reduction and measurements are performed using standard IRAF tasks and scripts.

The search is now focused on a sample of about 1000 objects,  mostly giant stars, many of them in the red giant clump region of the  HR diagram, and stars which have recently left the main sequence. The observing strategy has been optimized for the HET -- stars are randomly distributed over thy sky. %Measurements of a particular target star begin with 2--3 exposures, typically 3--6 month apart, to check for any RV variability exceeding a 30--50 m/s threshold. 
After almost five years of observations, more than one radial velocity measurement for $>800$ GK-giant stars with a 4-6 m/s precision was obtained. Currently, more than 50 planetary candidate companion stars are being monitored. The search is constrained by the position of the HET which allows to observe only targets  located on the northern hemisphere. A possibility of using similar instrument on the southern hemisphere gives an opportunity to extend the project.

\section{Southern African Large Telescope}
The Southern African Large Telescope (SALT), located in Sutherland, South Africa, is a telescope, which design is based on HET. Its primary spherical mirror is composed of 91 hexagonal mirrors, controlled with a Center of Curvature Alignment Sensor (CCAS), and has an effective diameter of 10~m. SALT can rotate only in the azimuth axis (fixed $37^o$ zenith angle). Hence its field of view has a specific shape. The maximum observing time for one object during single night varies from 45 to 150 minutes. During the observations, the image of a star is followed by a tracker positioned above the primary mirror. The tracker is equipped with spherical aberration corrector (SAC) and moves across the mirror on the focal plane. Although the fixed elevation angle of the telescope causes a limited observing window and varying effective surface of the mirror, this construction allows to avoid complicated control and optical systems. It also reduces a building costs of such a large instrument.

\section{SALT -- High Resolution Spectrograph}
The SALT-HRS will be one of the three instruments working with SALT, next to SALTICAM (science imager/photometer) and Robert Stobie Spectrograph (R=12 000).
The SALT--HRS, designed by The University of Canterbury, will be a  high resolution echelle spectrograph (R= 16500 - 65000) with a spectral range 370 -- 890 nm and allowing for $<10$ m s$^{-1}$ precision. The SALT-HRS is design as dual-beam fibre-fed echelle spectrograph. The cameras are all-refractive. The concept for SALT-HRS is to be an efficient single object spectrograph using pairs of 1.3-2.2'' optical fibers, one for star and one for background (sky). Therefore, with HRS, accurate background subtraction will be possible. Opportunity to obtain  spectra for more then one object at a time will is foreseen. However, it will limit the spectral range.
The spectrograph will be enclosed in a vacuum tank to improve stability of the instrument, and to ensure that the optics will remain in a perfect condition. It will be located underneath the telescope, in a temperature controlled enclosure.

\section{SALT Search for Planets  candidate star selection}
The selection of stars was based on their position on the HR diagram. Hence, the photometric and astrometric  data (position, parallax and proper motion) from (mainly) Tycho, Hipparchos~\cite{hip} and 2MASS~\cite{klei} catalogues were used to obtain effective temperatures (via effective temperature -- colour calibration) and absolute magnitudes. For this purpose an algorithm which would easily, quickly and efficiently select candidates for further research from available stars catalogues was developed.

To obtain effective temperature for each star, the temperature -- colour calibration for 6 colours by Ram\'irez and Mel\'endez~\cite*{rm2} was used, assuming solar metallicity for all stars. 
%%%% Fig 1 here
%\begin{center}
\begin{figure}[t]
\includegraphics[width=0.85\textwidth]{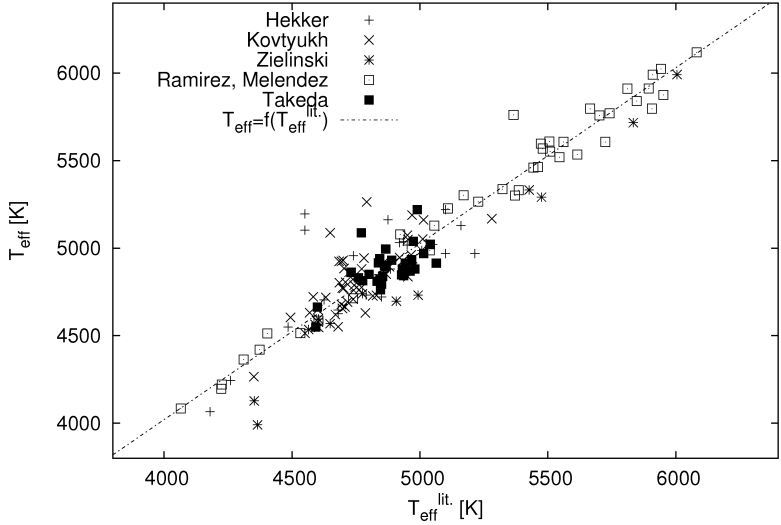}
\caption[]{Comparison between the effective temperature obtained in this work with effective temperature determinations collected from literature: Kovtyukh~\etal~\cite*{kov}; Hekker and Mel\'endez~\cite*{hek}; Takeda~\etal~\cite*{tak}; Zieli\'nski and Niedzielski~\cite*{ziel}; Ram\'irez and Mel\'endez~\cite*{rm1}.}
\end{figure}
%\end{center}
%%%%
Separate calibrations for different luminosity classes were applied. 

The initial sample of stars was  divided into two groups: giants and dwarfs. Two different methods were used: reduced proper motions -- $RPM=K+5.0\log_{10}\mu < 1$ for giants~\cite{nie} and empirical V vs. J, K, H relations for giants and dwarfs~\cite{jv}. 
For stars with unknown astrometric parallax, a spectrophotometric parallax (via Schmidt -- Kaler calibration) was used instead. It was calculated assuming that: (1)  there is a linear  relation between proper motion and the parallax (all stars in the sample have known proper motion, measured with good precision), what gives a initial estimated parallax value;
and (2) the difference between estimated parallax and spectroscopic parallax have a low influence on the effective temperature, especially for stars closer than 350 pc. 

The parallax was used to estimate extinction with {\it extinct.for} code by Hakkila~\etal~\cite*{hak}, which is based on interstellar gas and dust distribution in the Galaxy and requires both galactic coordinates and distance estimate of a star as  input parameters (see assumption 1).  Detailed description of the procedure and data for the complete sample of 270 000 stars are presented in Adam\'ow~\cite*{ja}.

The corrected parallax allowed us to determine the effective temperature and the absolute magnitude and thus place selected stars on the HR diagram. This information may be also useful to estimate stellar mass. The obtained temperatures were compared with results of other studies (Figure 1). The linear correlation coefficient for over 160 examined stars is 0.946 (the mean difference between  our and literature values of effective temperatures $\Delta T_{eff}$ is 96K), which proves that this method gives satisfactory results.
%
%%%%% Figure 3
%
\begin{figure}[t]
\centering
\includegraphics[width=0.85\textwidth]{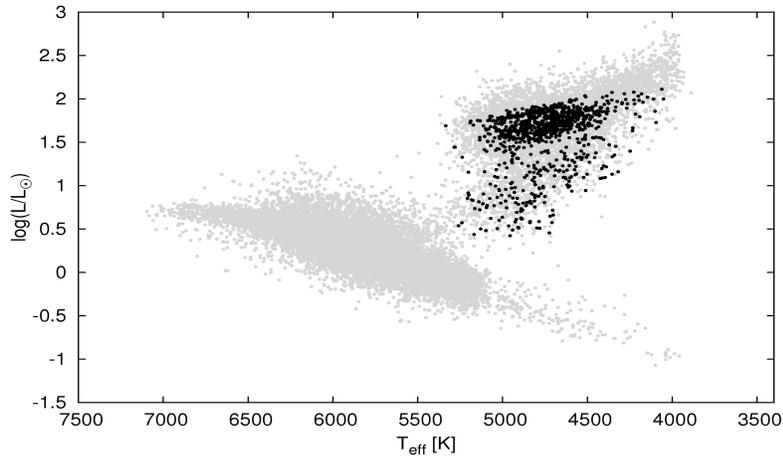}
\caption{Proposed SALT Planet Search targets (black circles )  on the HR diagram.}
\end{figure}
In this particular case study candidates for the future SALT Planet Search were selected according to the following criteria:
\begin{itemize}
\item position suitable for SALT (declination range from $-70^\circ$ to $+10^\circ$);
\item position on the HR diagram, which fulfils the project requirements (see Niedzielski \& Wolszczan 2008 for details);
\item visual magnitude dimmer than $8^m$ -- brighter stars are available to smaller telescopes;
\item the best temperature and absolute magnitude (or $L/L_\odot$) determination, which allows to place a star on the diagram more precisely.
\end{itemize}
Finally, more than 900 stars were selected and they are presented on Figure 2 as black circles. The grey circles represent all stars with highest precision temperature determination ($\sim$45 000 objects).  The selected  stars are  homogeneously distributed over the sky available for SALT what allows to use the queue scheduled mode in the most efficient way. 

The complete database with compiled literature data and information on estimated distances, reddening, luminosity  classes, temperatures and $\log_{10}L/L_{\odot}$ for more than 150 000 stars was created and may be used in future modifications of the program star selection. 

\section{Acknowledgements}
We acknowledge the financial support from the Polish Ministry of Science and 
Higher Education through grant 1P03D-007-30. The project was supported by UMK grant 413-A.
%
%%-----------------------------
%%      your bibliography
%%-----------------------------
%\bibliography{Adamow_m,Adamow_bib}

\end{document}